\documentclass[a4paper,11pt]{article}
\usepackage{pos}
\usepackage{comment}
\usepackage{slashed}
\usepackage{subfig}
\title{NLO QCD effects on angular observables in single Higgs production at electron-proton collider }
\ShortTitle{NLO QCD effects on angular observables at $ep$ collider}
\author*[a]{Pramod Sharma}
\author[b]{Biswajit Das}
\author[a]{Ambresh Shivaji}
\affiliation[a]{Department of Physical Sciences, Indian Institute of Science Education and Research Mohali, \\ 
Knowledge city, Sector 81, Manauli, PO, Sahibzada Ajit Singh Nagar, Punjab, India, 140306}

\affiliation[b]{The Institute of Mathematical Sciences,\\ 
	4th Cross St, CIT Campus, Taramani, Chennai, Tamil Nadu, India, 600113}

\emailAdd{ph19010@iisermohali.ac.in}
\emailAdd{biswajitd@imsc.res.in}
\emailAdd{ashivaji@iisermohali.ac.in}

\abstract{Properties of the Higgs boson ($H$) at current and future particle colliders are crucial to explore new physics beyond the standard model. In particular, experimental and theoretical outlooks at future colliders drive interest in Higgs to gauge boson couplings. Single Higgs production via vector-boson fusion allows probing Higgs couplings with massive vector bosons ($V = W, Z$). We consider electron-proton (eP) collider to study these couplings due to the low background. In a recent study, we considered the most general anomalous Higgs-vector boson ($HVV$) couplings and explored the potential of eP collider in constraining the parameters of $HVV$ couplings. Our results were based on leading order predictions in perturbation theory. We include further Next to Leading Order (NLO) corrections of Quantum Chromodynamic (QCD) in Standard Model signal to make precise predictions. In this talk, I will present the effect of NLO QCD corrections on the standard model and anomalous $HVV$ couplings.}

\FullConference{42nd International Conference on High Energy Physics (ICHEP2024)\\
18-24 July 2024\\
Prague, Czech Republic\\}
\begin{document}
\maketitle
\section{Introduction}
The scalar boson with a mass of 125 GeV, discovered by the ATLAS and CMS collaborations at the Large Hadron Collider (LHC) \cite{ATLAS:2012yve,CMS:2012qbp}, is consistent with the Standard Model (SM) Higgs boson. The SM has demonstrated remarkable success in explaining a wide range of experimental measurements. The investigation of electroweak symmetry breaking via the Higgs-to-vector boson couplings ($HVV$) is important for new physics searches at current and future colliders. The $HVV$ couplings in the Kappa framework have been measured with an accuracy of 7-8\% by CMS \cite{CMS:2022dwd} and ATLAS \cite{ATLAS:2022vkf} using Run-II LHC data. The new physics allowed by such a measurement can be parametrized by the most general Lorentz structure of two rank tensors made of momenta of vector bosons. The expression for the $HVV$ vertex is given by,
\begin{align}
	\Gamma_{HVV}^{\mu\nu}(p_1,p_2) = & g_V m_V \kappa_V g^{\mu\nu} - \frac{g}{m_W} [\lambda_{1V} (p_1^\nu p_2^\mu - g^{\mu \nu} p_1.p_2) \notag \\
& + \lambda_{2V} (p_1^{\mu}p_1^{\nu}+p_2^{\mu}p_2^{\nu}-g^{\mu \nu} p_1.p_1-g^{\mu \nu} p_2.p_2) \notag \\
& -  \widetilde{\lambda}_V ~ \epsilon^{\mu \nu \alpha \beta} p_{1 \alpha} p_{2 \beta} ].
	\label{Eq:BSMcoup}
\end{align}
Here, $p_1$ and $p_2$ are the momenta of the vector bosons. The BSM parameters $\kappa_V$ and $\lambda_{1V, 2V}$ are associated with CP-even couplings, while $\widetilde{\lambda}_V$ is associated with CP-odd coupling.

We consider electron-proton collider to probe the $HZZ$ coupling in the neutral current (NC) process, $e^- p \rightarrow e^- H j + X$, while the $HWW$ coupling in the charged current (CC) process, $e^- p \rightarrow \nu_e H j + X$. In a previous study \cite{Sharma:2022epc}, we analyzed the sensitivity of these couplings using the absolute value of azimuthal correlation ($|\Delta \phi|$) between two final state particles. Distinct Lorentz structures of the CP-even and CP-odd components of these couplings motivate us to explore new angular observables that could be sensitive to the individual components. Since the CC process has missing energy in the final state, whereas the NC process has an $e^-$ in the final state, the latter provides flexibility to study various angular observables. In this letter, we present the sensitivity of angular observables to the $HZZ$ coupling and constraints on the parameters associated with this coupling. This analysis is performed at the tree level in perturbation theory. To minimize theoretical uncertainties, we further incorporate next-to-leading (NLO) corrections to CC and NC processes in QCD. Previous studies \cite{Blumlein:1992eh,Jager:2010zm} have addressed NLO QCD corrections for CC and NC processes in the SM. In our work, we show the effect of NLO QCD correction on angular observables sensitive to BSM physics.

\section{Angular Observables}
In this work, we follow the parton level analysis performed in Ref.~\cite{Sharma:2022epc}, where the signal and background processes are discussed in detail. Events are generated using \texttt{MadGraph}, with the electron and proton beam energies set to 60 GeV and 7 TeV, respectively. In the signal process, the decay of the Higgs boson is considered in its dominant decay channel, $H \rightarrow b\bar{b}$.  The signal is identified with $e^-$, two b-jets, and one light jet, $j = u,c,d,s,g$.
\begin{figure}[htp]
	\centering
	\subfloat[]{\includegraphics[width=0.35 \textwidth]{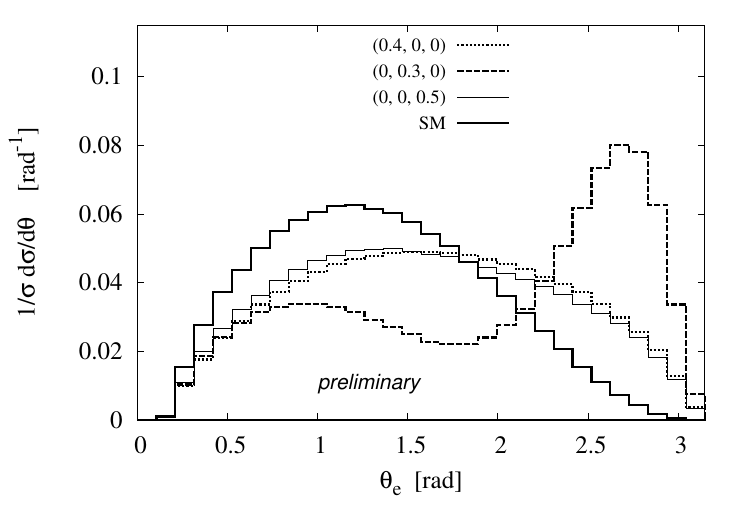} \label{thetaFig}}
	\subfloat[]{\includegraphics[width=0.35 \textwidth]{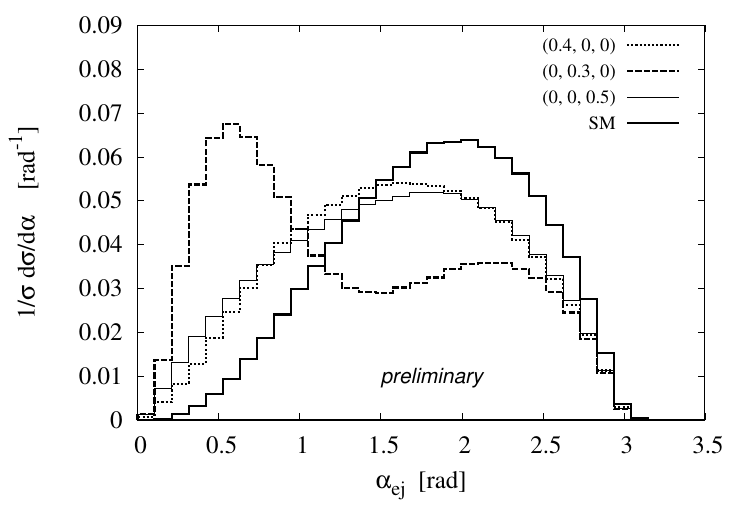} \label{alphaFig}}\\
    \subfloat[]{\includegraphics[width=0.35 \textwidth]{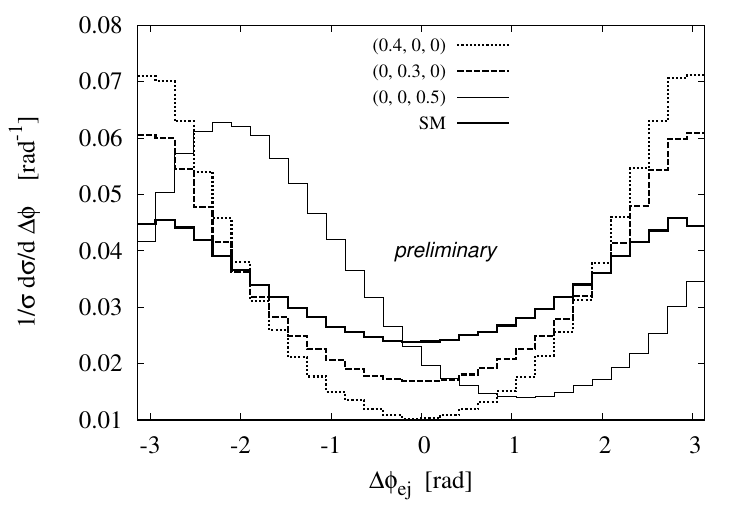} \label{dphiFig}}
    \subfloat[]{\includegraphics[width=0.35 \textwidth]{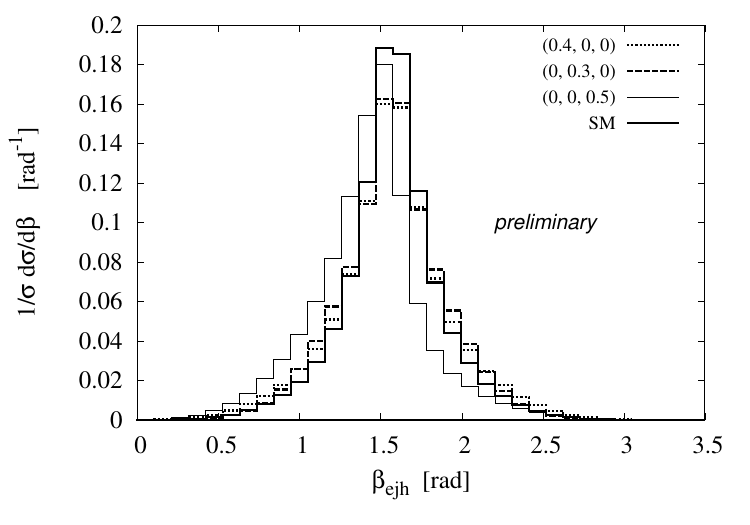} \label{betaFig}}
	\caption{Angular observables for SM and banchmark values of BSM parameters ($\lambda_{1Z}, \lambda_{2Z}$, $\widetilde{\lambda}_Z$) in the NC process.}
	\label{obs}
\end{figure}

The angular distributions described in Eq.~\ref{obs_eq}, defined in the lab frame, are as follows: $\theta_e$ represents the angle between the scattered electron ($e^-$) and the incident electron beam direction. The angle $\alpha_{ej}$ is the angle formed between the $e^-$ and the $jet$. 
Lastly, $\beta_{ejh}$ denotes the angle between the Higgs boson and the normal to the plane defined by the $e^-$ and the $jet$. 

	\begin{align}
	& \theta_{e} = \text{cos}^{-1}\left( \frac{\vec{p}_e.\hat{z}}{|\vec{p}_e|} \right), \hspace{4cm}
		 \alpha_{ej} = \text{cos}^{-1}\left( \frac{\vec{p}_e . \vec{p}_j}{|\vec{p}_e||\vec{p}_j|} \right) \notag \\
	&	\Delta \phi_{ej} = \text{tan2}^{-1}\left( \frac{p_{ey}}{p_{ex}} \right) - \text{tan2}^{-1}\left( \frac{p_{jy}}{p_{jx}} \right), \hspace{1cm} 
		\beta_{ejh} = \text{cos}^{-1}\left( \frac{(\vec{p}_e \times \vec{p}_j).\vec{p}_h}{|\vec{p}_e \times \vec{p}_j| |\vec{p}_h|} \right)	
	\label{obs_eq}
\end{align}
Here, $\vec{p}_i~ (i = e, j, h)$ are the momenta of final state particles: $e^-$, Higgs, and $jet$ in the NC process. Fig. \ref{obs} shows the sensitivity of angular observables to individual BSM coupling. For demonstration, we use the benchmark values $\lambda_{1Z} = 0.4$, $\lambda_{2Z} = 0.3$, and $\widetilde{\lambda}_Z = 0.5$ of these couplings. As shown in Fig.~\ref{thetaFig}, the polar angle $\theta_e$ is most sensitive to $\lambda_{2Z}$ among all BSM couplings. We notice in Fig.~\ref{alphaFig} that the $\alpha_{ej}$ distribution efficiently distinguishes the $\lambda_{2Z}$ coupling from the SM, while this distribution shows behavior similar to the SM for the other couplings, $\lambda_{1Z}$ and $\widetilde{\lambda}_Z$. The  $\Delta \phi_{ej}$ distribution, as given in Eq.~\ref{obs_eq}, is defined using the tan2$^{-1}$ (quadrant inverse tangent) function to consider all possible directions in the transverse plane of the detector. We notice in Fig.~\ref{dphiFig} that the $\Delta \phi_{ej}$ distribution is most sensitive to $\widetilde{\lambda}_Z$. As shown in Fig.~\ref{betaFig}, the $\beta_{ejh}$ distribution is shifted towards left about $\beta_{ejh}= \pi/2$ for $\widetilde{\lambda}_Z$ hence providing better sensitivity to this coupling.

\begin{figure}[htp]
	\centering
	\includegraphics[width=0.4 \textwidth]{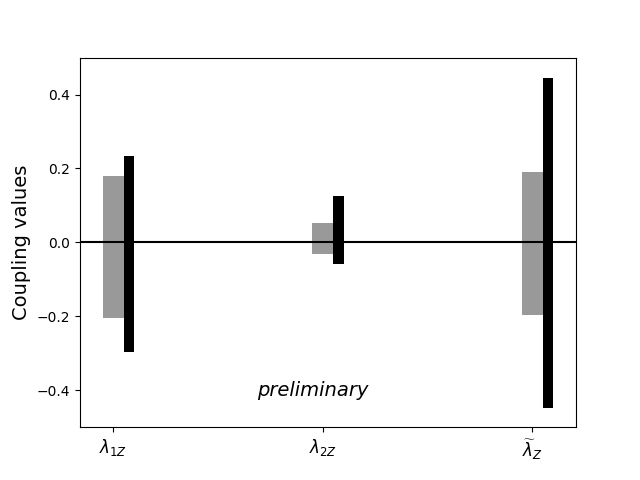}
	\caption{Constraints on angular observables from this study (grey bars) and from $|\Delta \phi_{ej}|$ distrbution (black bars) of Ref. \cite{Sharma:2022epc} which was based on a $\chi^2$ analysis with two bins in the distribution. Constraints are obtained at 0.1 $ab^{-1}$ luminosity and 2$\sigma$ confidence level.}
    \label{bounds}
\end{figure}

We give constraints on BSM parameters based on the most sensitive angular observables. The $\chi^2$ analysis is used by taking bins size of 9$^{\circ}$ in the distributions to estimate constraints on individual parameters.
Fig. \ref{bounds} presents the constraints on these couplings at a luminosity of 0.1 $ab^{-1}$ and 2$\sigma$ confidence level. The $\theta_{e}$ distribution is most sensitive to $\lambda_{2Z}$, while the $\Delta \phi_{ej}$ and $\beta_{ejh}$ distributions are most sensitive to $\widetilde{\lambda}_Z$, we give constraints on $\lambda_{2Z}$ and $\widetilde{\lambda}_Z$ using the $\theta_{e}$ and $\Delta \phi_{ej}$ (or $\beta_{ejh}$) distributions, respectively. We compare these constraints with those from our previous analysis based on $|\Delta \phi_{ej}|$. We find that using these new angular observables, the constraints on $\lambda_{2Z}$ and $\widetilde{\lambda}_Z$ are narrowed down by $\sim$57\%  compared to the constraints obtained from $|\Delta \phi_{ej}|$ in Ref. \cite{Sharma:2022epc}. Further, we note that $|\Delta \phi_{ej}|$ is still the best observable to constrain $\lambda_{1Z}$.

\section{NLO QCD Corrections}
We are interested in studying NLO QCD corrections on both NC and CC processes. Contribution at NLO requires virtual and real-emission corrections to the tree-level process, $e^- (k_1)~ q(p_b) \rightarrow e^-/\nu_e(k_2)~ H(p_H)$ $q\prime (p_1)$. There is only one tree-level diagram and one virtual diagram at NLO. There are two sub-processes for the real emission diagrams and each subprocess has two real emission diagrams. The sub-processes are
\begin{align}
	e^-(k_1)~ q(p_b)  \rightarrow e^-/\nu_e(k_2)~ q\prime(p_1)~ g(p_2)~ H(p_H) \\
	e^-(k_1)~ g(p_b)  \rightarrow e^-/\nu_e(k_2)~ q(p_1)~ \bar{q\prime}(p_2)~ H(p_H).
\end{align}
 
The virtual diagram is UV and IR divergent and we can write the exact expression for the virtual diagram in the t'Hooft-Veltman (HV) regularization scheme as

\begin{equation}
	\quad\mathcal{M}^V=\frac{\alpha_s(\mu_R)}{2\pi}.\frac{(4\pi)^\epsilon}{\Gamma(1-\epsilon)} .C_F.\Big(\frac{\mu^2}{t}\Big)^\epsilon.\Big\{-\frac{1}{\epsilon^2}-\frac{3}{2\epsilon}-4+\mathcal{O}(\epsilon)\Big\}\times \mathcal{M}^B,
	\nonumber
\end{equation}
where $\mathcal{M}^B$ is the born level amplitude. Here $\epsilon$ is defined as $\epsilon=(4-d)/2$, where $d$ is the space-time dimension. $\alpha(\mu_R)$ is the strong coupling at renormalisation scale $\mu_R$, and the color factor $C_F = 3/4$. 
The real emission diagrams are also divergent in the soft and collinear regions. The virtual and real amplitudes are separately divergent but their sum is finite. We adopt the Catani-Seymour dipole subtraction scheme following the Ref.~\cite{Catani:1996vz} to handle the IR singularities. Following the reference, the NLO cross-section is written as,
	\begin{equation}
	\begin{aligned}
	 \sigma^{\rm NLO}=\int_{m+1}\big[d\sigma^R-\sum_{\text{dipoles}}d\sigma^B\otimes dV_{\text{dipole}}\big]_{\epsilon=0}+\int_m \big[d\sigma^V+d\sigma^B\otimes \mathbf{I}\big]_{\epsilon=0}\\
	\end{aligned}
	\label{equ_nlo}
\end{equation}

Here, $\sigma^B$, $d\sigma^V$, and $\sigma^R$ are the Born, virtual, and real emission cross-sections. The $\bf{I}$ term cancels all singularities in the virtual amplitude which has been regularised in $d$-dimention. The dipoles have the same point-like singular behavior as the real amplitudes in the soft and collinear regions.
The two integrals in Eq.~\ref{equ_nlo} are finite in $4$-dimension and one can perform Monte-Carlo integration separately.
For this process, the insertion operator $\mathbf{I}$ can be written as

\begin{equation}
	\begin{aligned}
		\mathbf{I}(\varepsilon)=\frac{\alpha_s(\mu_R)}{2\pi}.\frac{(4\pi)^\epsilon}{\Gamma(1-\epsilon)} .2C_F.\Big(\frac{\mu^2}{t}\Big)^\epsilon.\Big\{\frac{1}{\epsilon^2}+\frac{3}{2\epsilon}+5-\frac{\pi^2}{2}+\mathcal{O}(\epsilon)\Big\}.\nonumber
	\end{aligned}
\end{equation}
 We have computed the dipoles for this process following the Ref.~\cite{Catani:1996vz} and they show the exactly same behavior.
To check the reliability of the calculation, we have compared the tree-level and real emission amplitude with \texttt{MadGraph} for a few sets of phase space points. We also compared the total cross section for leading order processes with \texttt{MadGraph}. All the checks are in agreement.

\begin{figure}[htp]
	\centering
     \includegraphics[width=0.37 \textwidth]{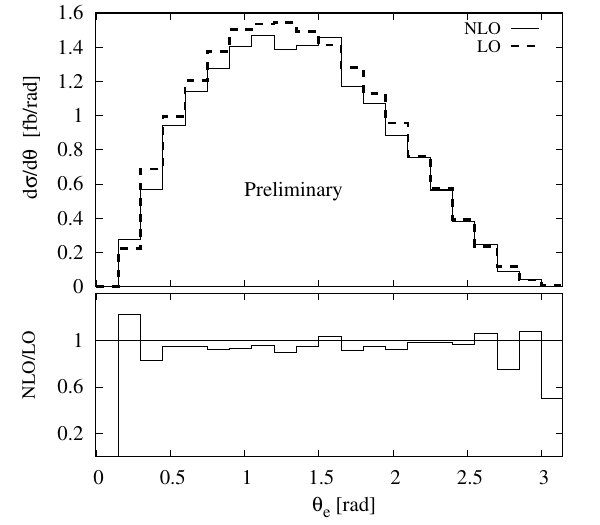} 
	 \includegraphics[width=0.37 \textwidth]{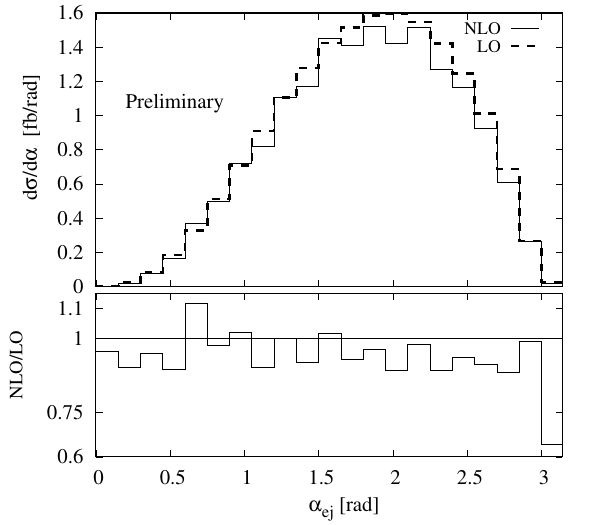}\\
	\includegraphics[width=0.37 \textwidth]{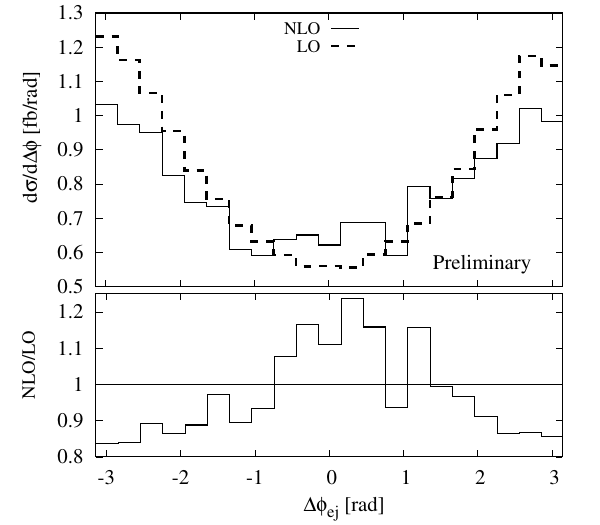} 
	\includegraphics[width=0.37 \textwidth]{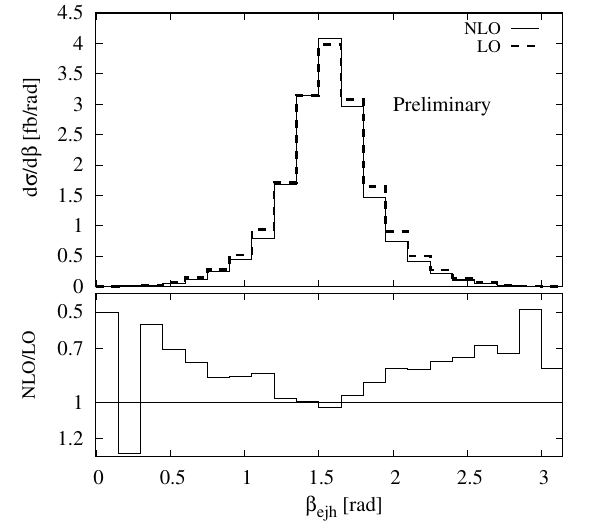}\\
	\caption{Angular distributions at LO and NLO for the NC process, $e^-p\rightarrow e^- Hj$.}
	\label{NLO}
\end{figure}

	We use the \texttt{CT18LO} PDF set for LO calculations and \texttt{CT18NLO} for NLO calculations. The values of the SM parameters are as follows: the weak boson masses, $M_W = 80.3692$ GeV and $M_Z = 91.118$ GeV. Quark and lepton are taken as massless. The evolution of PDF and running of strong coupling depends on the renormalization ($\mu_R$) and factorization ($\mu_F$) scales. We set these scales equal to a dynamical scale as
	\begin{equation}
\mu_R=\mu_F=\mu_0=\frac{1}{3}\Big(p_{T,l}+\sqrt{p_{T,H}^2+M^2_{H}}+p_{T,j}\Big).\nonumber
\end{equation}
We consider electon-proton collision with electron beam energy ($E_e$) = 60 GeV and proton beam energy ($E_p$) = 7 TeV. The total cross-section of NC and CC processes at LO ($\sigma_0$) and NLO ($\sigma_{qcd}^{NLO}$) are shown in the table given below.
\begin{center}
	\begin{tabular}{ |c |c|c|c|} 
		\hline
		Process& $\sigma_0$ (fb) & $\sigma^{NLO}_{qcd}$ (fb) &$RE$ (\%)\\ 
		\hline
		$e^-p\rightarrow$$e^-Hj$&$17.44$ &$18.25$  & $4.6$ \\
		\hline
		$e^-p\rightarrow$$\nu_eHj$&91.86&96.55 &$5.1$\\
		\hline
	\end{tabular}
	\label{Xsec}
\end{center}
The enhancement in the cross-section relative to LO, defined by RE, is 4.6\% and 5.1\% for the NC and CC processes, respectively. Since the diagrams involved in the QCD corrections for NC and CC processes are similar, the corrections are also similar in size. Fig. \ref{NLO} demonstrates the effect of NLO corrections on the angular observables $\theta_e$, $\alpha_{ej}$, $\Delta \phi_{ej}$, and $\beta_{ejh}$ distributions in the NC process of the SM in Higgs + 1 jet events. We observe that the $\theta_e$ and $\alpha_{ej}$ distributions exhibit flat corrections in the mid-region, while $\Delta \phi_{ej}$ and $\beta_{ejh}$ show nontrivial corrections from QCD. This suggests that QCD corrections may be important for constraints on $\widetilde{\lambda}_Z$.

\section{Conclusion}
We show the potential of angular observables to probe the $HZZ$ coupling at $ep$ collider with a center of mass energy of $\sim$1.3 TeV. The angular observables, $\alpha_{ej}$ and $\Delta \phi_{ej}$, put $\sim$57\% stronger bounds on $\lambda_{2Z}$ and $\widetilde{\lambda}_Z$ as comapred to the $|\Delta \phi_{ej}|$ distribution. To improve precision, we calculate the QCD NLO corrections to both NC and CC processes. At NLO, enhancements of 4.6\% and 5.1\% are noted with respect to LO in the NC and CC processes. We also show the effects of NLO QCD corrections on the angular observables considered in this study.

\end{document}